\documentclass[aps,prl,twocolumn,tightenlines,amsmath,amssymb,superscriptaddress]{revtex4-1}

\usepackage{graphicx}
\usepackage{amsmath}
\usepackage{amssymb}
\usepackage{epstopdf}
\usepackage{color}
\usepackage{lipsum}
\usepackage{hyperref}
\usepackage{ulem}
\usepackage[dvipsnames]{xcolor}

\DeclareGraphicsExtensions{.eps}

\newcommand{\bra}[1] {\langle #1|}
\newcommand{\ket}[1] {|#1 \rangle}

\newcommand{\tket}[1]{$|#1\rangle$}

\newcommand*{\rom}[1]{\expandafter\@slowromancap\romannumeral #1@}

\begin{document}

\title{Generation of non-classical light in a photon-number superposition}
    
	\author{J.~C.~Loredo}
\thanks{Equally contributing authors\\juan.loredo@u-psud.fr\\carlos.anton-solanas@c2n.upsaclay.fr}
\affiliation{CNRS Centre for Nanoscience and Nanotechnology, Universit\'e Paris-Sud, Universit\'e Paris-Saclay, 91120 Palaiseau, France}
	\author{C.~Ant\'on}	
\thanks{Equally contributing authors\\juan.loredo@u-psud.fr\\carlos.anton-solanas@c2n.upsaclay.fr}
\affiliation{CNRS Centre for Nanoscience and Nanotechnology, Universit\'e Paris-Sud, Universit\'e Paris-Saclay, 91120 Palaiseau, France}
	\author{B.~Reznychenko}
\affiliation{{Univ. Grenoble Alpes, CNRS, Grenoble INP, Institut N\'eel, 38000 Grenoble, France}}
\author{P.~Hilaire}
	\affiliation{CNRS Centre for Nanoscience and Nanotechnology, Universit\'e Paris-Sud, Universit\'e Paris-Saclay, 91120 Palaiseau, France}
\author{A.~Harouri}
	\affiliation{CNRS Centre for Nanoscience and Nanotechnology, Universit\'e Paris-Sud, Universit\'e Paris-Saclay, 91120 Palaiseau, France}
\author{C.~Millet}
	\affiliation{CNRS Centre for Nanoscience and Nanotechnology, Universit\'e Paris-Sud, Universit\'e Paris-Saclay, 91120 Palaiseau, France}
\author{H.~Ollivier}
	\affiliation{CNRS Centre for Nanoscience and Nanotechnology, Universit\'e Paris-Sud, Universit\'e Paris-Saclay, 91120 Palaiseau, France}
\author{N.~Somaschi}
	\affiliation{Quandela SAS, 86 rue de Paris, 91400 Orsay, France}
\author{L.~De~Santis}
	\affiliation{CNRS Centre for Nanoscience and Nanotechnology, Universit\'e Paris-Sud, Universit\'e Paris-Saclay, 91120 Palaiseau, France}
\author{A. Lema\^itre}
	\affiliation{CNRS Centre for Nanoscience and Nanotechnology, Universit\'e Paris-Sud, Universit\'e Paris-Saclay, 91120 Palaiseau, France}
\author{I.~Sagnes}
	\affiliation{CNRS Centre for Nanoscience and Nanotechnology, Universit\'e Paris-Sud, Universit\'e Paris-Saclay, 91120 Palaiseau, France}
\author{L.~Lanco}
	\affiliation{CNRS Centre for Nanoscience and Nanotechnology, Universit\'e Paris-Sud, Universit\'e Paris-Saclay, 91120 Palaiseau, France}
	\affiliation{Universit\'e Paris Diderot, Paris 7, 75205 Paris CEDEX 13, France}
\author{{A.~Auff\`eves}}
	\affiliation{{Univ. Grenoble Alpes, CNRS, Grenoble INP, Institut N\'eel, 38000 Grenoble, France}}
\author{O.~Krebs}
	\affiliation{CNRS Centre for Nanoscience and Nanotechnology, Universit\'e Paris-Sud, Universit\'e Paris-Saclay, 91120 Palaiseau, France}
\author{P.~Senellart}
	\email{pascale.senellart-mardon@c2n.upsaclay.fr}
	\affiliation{CNRS Centre for Nanoscience and Nanotechnology, Universit\'e Paris-Sud, Universit\'e Paris-Saclay, 91120 Palaiseau, France}

\begin{abstract}
{The ability to generate light in a pure quantum state is essential for advances in optical quantum technologies. However, obtaining quantum states with control in the photon-number has remained elusive. Optical light fields with zero and one photon can be produced by single atoms, but so far it has been limited to generating incoherent mixtures, or coherent superpositions with a very small one-photon term. Here, we report on the on-demand generation of quantum superpositions of zero, one, and two photons via pulsed coherent control of a single artificial atom. Driving the system up to full atomic inversion leads to the generation of quantum superpositions of vacuum and one photon, with their relative populations controlled by the driving laser intensity. A stronger driving of the system, with $2\pi$-pulses, results in a coherent superposition of vacuum, one, and two photons, with the two-photon term exceeding the one-photon component, a state allowing phase super-resolving interferometry. Our results open new paths for optical quantum technologies with access to the photon-number degree-of-freedom.}
\end{abstract}

\maketitle

Controlling the photon-number in a light pulse has been a primary task enabling progress in {optical} quantum technologies~\cite{OBrien:2009rm,Erhard:2018fk}. Single-, and N-photon sources~\cite{NPhoton:EWaks,MPQI:Pan,10photEnt:Pan} are at the heart of future quantum communication networks~\cite{SecureQKD:Scarani,QKDMetro:18}, sensors~\cite{Brida:2010fk,Schwartz:2013qf}, as well as optical quantum computers~\cite{KLM,OBrienOptQC} and simulators~\cite{Lanyon:2010bh,EQS:Loredo,EQS:JWP,Santagatieaap9646}. These achievements make use of the interference of indistinguishable single-photons, allowing the realisation of quantum gates~\cite{OBrien:2003cs,Patele1501531}, and protocols such as quantum teleportation~\cite{Wang:2015mz} and entanglement swapping~\cite{EntSwap:15}. The one-photon term has been exploited heretofore, and the vacuum component has been considered detrimental to the overall protocol efficiency, motivating a quest for deterministic sources producing single-photon Fock states with no vacuum component~\cite{somaschi2016,nearIdealSPS:Pan,senellart2017,IndPhoton:Lodahl17}---a challenging task, to say the least. If vacuum is set instead in a quantum superposition with the single-photon, one could use it to encode quantum information in the photon-number---becoming a resource for optical quantum information processing. For instance, vacuum within a pure quantum state can be exploited in quantum teleportation~\cite{vacuumTel:Lombardi}, or quantum random number generators~\cite{Gabriel:2010lq}. However, obtaining quantum superpositions in the photon-number basis has so far demanded complex quantum state engineering and conditioned state preparation~\cite{Bimbard:2010yu,Fuwa:2015gf}.
 
The text-book model of a quantum emitter is a two level atom---a system shown to generate quantum light  in various excitation regimes. Incoherent non-resonant excitation of natural~\cite{Aspect:1986} and artificial atoms~\cite{lounis2000,michlerLowT,michlerRoomT} can produce {optical} fields with a large single-photon component, but without coherence in the photon-number basis due to the incoherent creation process of the atomic population. In contrast, coherent driving of an atom can in principle be used to transfer the coherence between the atomic ground and excited state  to the emitted light field. This has so far been explored in the weak-excitation regime to produce quantum light that exhibits coherence with the driving laser---observed with atoms~\cite{jessen1992}, as well as semiconductor quantum dots~\cite{nguyen2011,matthiesen2012,atature2013,proux2015}. This regime has been shown to produce squeezed light where  an atomic dipole---with vanishing population---elastically scatters a coherent superposition of vacuum and a small one-photon term~\cite{schulte2015}. Generating a photon-number superposition with large single-photon population requires to create an atomic population---inherently coupled to its environment---that remains insensitive to any decoherence until spontaneous emission takes place. To the best of our knowledge, the  generation of photon-number quantum superpositions under strong coherent driving has not been reported so far, neither with natural atoms, nor with artificial ones.

In this work, we report on the on-demand generation of quantum superpositions in the photon-number basis, in light pulses emitted by a single artificial atom. We observe superpositions of zero, one, and two photons emitted from semiconductor quantum dots coupled to optical microcavities~\cite{somaschi2016,giesz2016}. We use pulsed coherent driving, beyond full inversion of the atomic population, and perform interferometric measurements with a path-unbalanced Mach-Zehnder interferometer (MZI). As supported by our theoretical calculations, phase-dependent oscillations at the interferometer output demonstrate the production of coherent superpositions of vacuum, one, and two photons. Below $\pi$-pulse driving, we obtain superpositions of vacuum and one-photon Fock states, with their relative populations controlled by the driving laser intensity. By driving the quantum dot with $2\pi$-pulses, we obtain a state with the two-photon component larger than the one-photon population---a state allowing phase super-resolving interferometry, and incidentally resembling a small Schr\"odinger-cat state.

	\begin{figure}[b!]
		\centering
		\includegraphics[width=.5\textwidth]{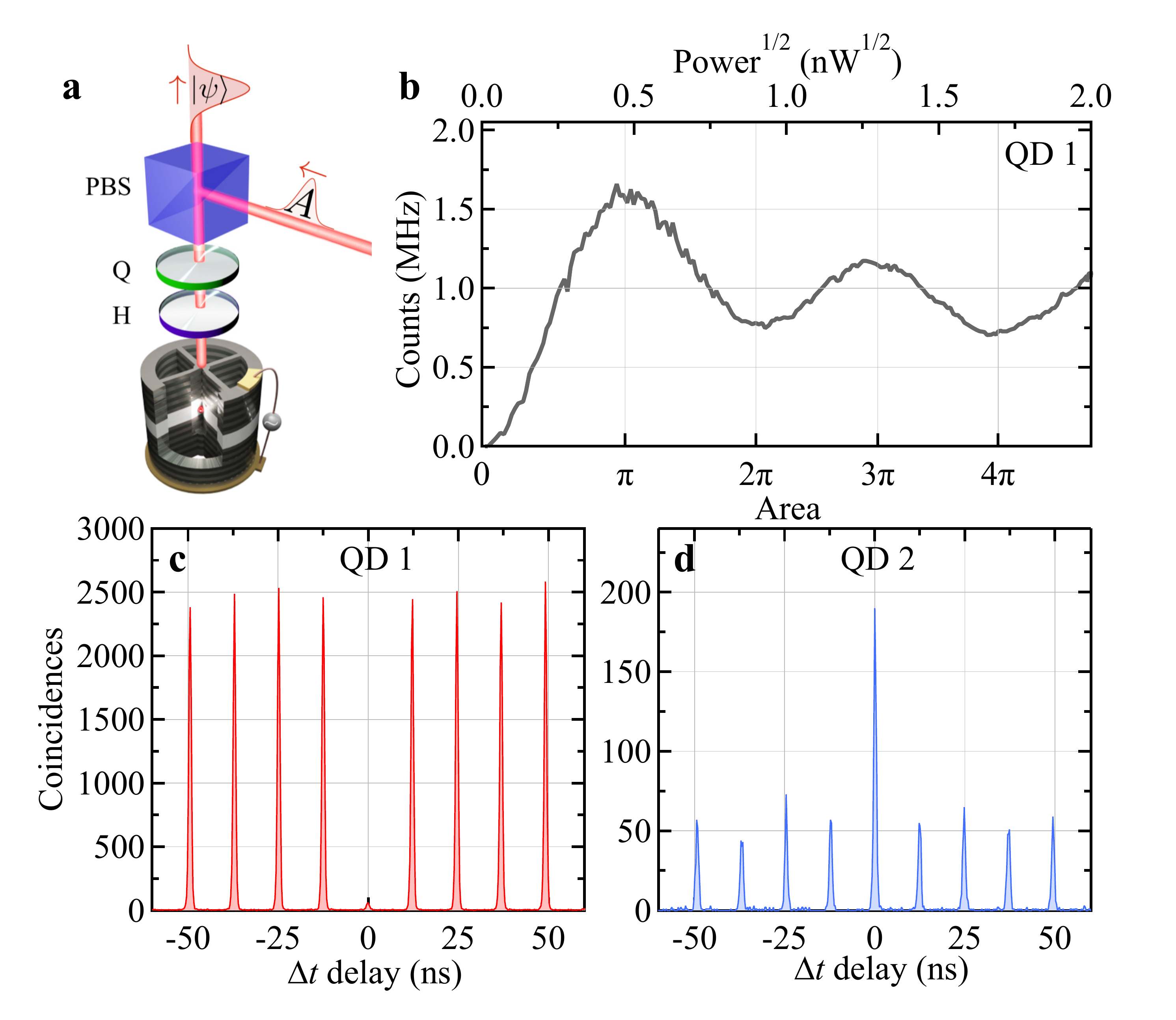}\vspace{-3mm}	
		\caption{{\bf Coherent control of an artificial atom}. {\bf a}  Schematics of the setup. A single semiconductor QD is kept in a cryostat at 9 K, and is excited under pulsed resonant excitation. The QD emitted state $| \Psi \rangle$ is separated from the laser in a cross-polarisation scheme, by using a polarising beamsplitter (PBS), a quarter- (Q), and half-wave plate (H). {\bf b} Rabi oscillation of the QD coherent driving. The emission is collected by a single-mode fibre and directly detected with an APD. {\bf c} Second-order autocorrelation function $g^{(2)}(\Delta t)$ measured at $\pi$-pulse with QD1, and {\bf d} at $2\pi$-pulse driving with QD2.}
	\label{fig:1}
	\end{figure}
\noindent{\bf Coherent driving and photon statistics~\\}
We investigate semiconductor devices consisting of a single quantum dot (QD) positioned with nanometer-scale accuracy at the centre of a connected-pillar cavity \cite{dousse2008,nowak2014,somaschi2016}. The QD layer is inserted in a p-i-n diode structure, and electrical contacts are defined to control the QD resonance through the confined Stark effect. We note that the experimental results reported here have been observed on various QD-cavity devices. We focus hereafter on two devices: a neutral (QD1) and a charged (QD2) exciton coupled to the cavity mode, {see Methods}. QD1 (QD2) is excited resonantly with linearly-polarised 40~ps (15~ps) laser pulses at $925$~nm, and {its} emission is collected using a crossed-polarisation scheme that separates it from the laser, see Fig.~\ref{fig:1}a. Figure~\ref{fig:1}b shows the detected countrates for QD1 as a function of the excitation pulse area $A$, evidencing well defined Rabi oscillations. The signal is damped by spontaneous emission due  to the relatively long 40~ps excitation pulses~\cite{giesz2016} {as compared to the measured emission decay time of $166{\pm}16$~ps}. Second-order autocorrelation functions $g^{(2)}(\Delta t)$ measured along Rabi cycles evidence a distinct and complementary behaviour between $\pi$-, and $2\pi$-pulse driving. A pronounced antibunched photon statistics at $\pi$-pulse {is observed for both QD1 and QD2}, with {$g^{(2)}_{\pi}(0){=}0.037{\pm}0.002$} {for QD1, see Fig.~\ref{fig:1}c. Such observations show light wavepackets consisting mostly of either vacuum or one photon. {However, bunched statistics is observed at $2\pi$-pulse for QD2, with $g^{(2)}_{2\pi}(0){=}2.98{\pm}0.11$}, see Fig.~\ref{fig:1}d. This evidences, as recently observed~\cite{Fischer:2017zr}, wavepackets containing two-photon populations. In the following, we investigate the nature of light in the photon-number degree-of-freedom: whether it contains photon Fock-states emitted in a mixture or in a pure quantum state.

	\begin{figure*}[htp!]
		\centering
		\includegraphics[width=.95\textwidth]{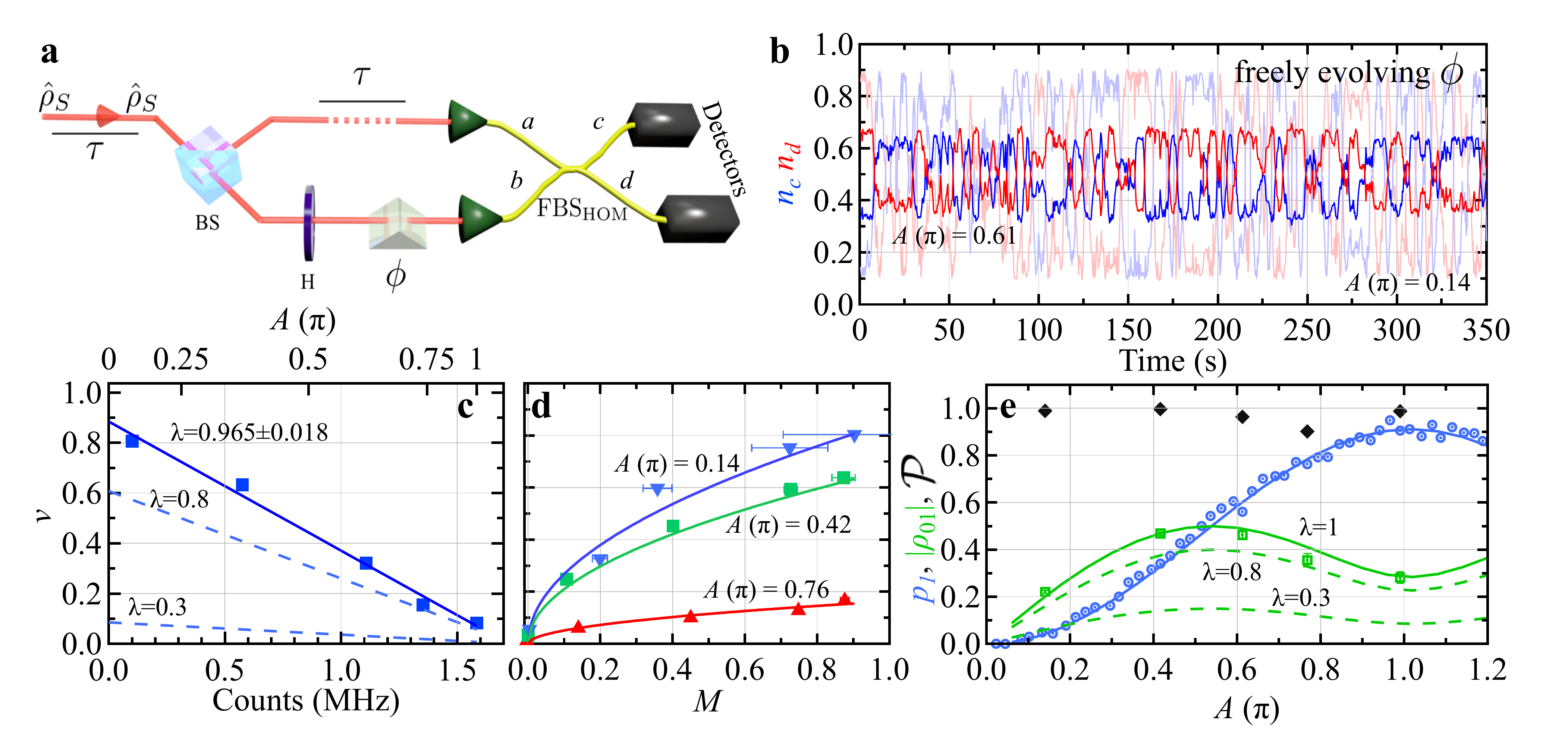}\vspace{-3mm}	
		\caption{{\bf Quantum superposition of vacuum and one photon.} {\bf a} Sketch of the MZI used to probe coherences in the photon-number. The MZI delays one arm by $\tau{=}12.34$~ns as to allow interference of two consecutive wave-packets in the fibre beamsplitter $\textrm{FBS}_\textrm{HOM}$. The phase $\phi$ between the two arms of the MZI is not stabilised and thus it freely evolves in time. A half-wave plate H in one arm tunes the photon distinguishability via their polarisation.  {\bf b} Normalised single countrates $n_c$ (blue) and $n_d$ (red) for a pulse area $A{=}0.61\pi$. Light blue (light red) traces display $n_c$ ($n_d$) for $A{=}0.14\pi$. Each data point here was accumulated for ${\sim}300$~ms. {\bf c} Measured visibility $v$ (blue squares) as a function of the countrates detected from our first collecting fibre. The blue solid line is a linear fit used to obtain the purity of the generated state, and the dashed blue lines consider lower purity values. {\bf d} Visibility $v$ in terms of the photon indistinguishability $M$ (varied via polarisation). Blue, green, and red data points are taken for pulse areas $A$ of $0.14\pi$, $0.42\pi$, $0.76\pi$, respectively; and their corresponding curves follow the theoretical model $v{=}\lambda^2p_0\sqrt{M}$. {\bf e} Blue line: theoretical prediction for the probability $p_1$ of the QD to emit one photon. Blue data points: experimental one-photon population. Green full line: theoretical prediction of the photon-number coherence amplitude ($|\rho_{01}|{=}\lambda \sqrt{p_0 p_1}$) assuming that the emitted state is pure ($\lambda{=}1$). Dashed green lines same: same as before for cases with less purity. Green data points: extracted values for $|\rho_{01}|$ deduced from the measured visibilities. Black data points: extracted values of the purity $\mathcal{P}$.}
	\label{fig:2}
	\end{figure*}	
\noindent{\bf Quantum superposition of zero and one photon~\\}
The Hong-Ou-Mandel (HOM) effect~\cite{HOM87} describes two single-photons {simultaneously impinging} on a beamsplitter. If the photons are polarisation, spatially, and frequency indistinguishable,  they bunch at the output of the beamsplitter{---a behaviour exclusively of quantum mechanical origin. This requires that the interfering photons are in the same pure quantum state in these degrees-of-freedom.

As discussed now, interference can also be used to unravel coherences in the Fock-state basis. Consider a beamsplitter with inputs $a,b$, and outputs $c,d$, onto which pure states of photon-number superpositions impinge. These are in the form} {$|\Psi_a\rangle{=}\sqrt{p_0}|0_a \rangle{+}\sqrt{p_1}e^{i \alpha}|1_a \rangle$, and $|\Psi_b\rangle{=}\sqrt{p_0}|0_b \rangle{+}\sqrt{p_1}e^{i \left(\alpha{+}\phi\right)}|1_b \rangle$}, with $p_0{+}p_1{=}1$, $p_{0,1}$ the vacuum and one-photon populations, and $\phi$ a relative phase between the states. When ${p_1}{=}1$, their quantum interference leads to the well known two-photon output state $\left(|2_c0_d\rangle{-}|0_c2_d\rangle\right ){/}\sqrt{2}$---the HOM effect. However, as soon as ${p_1}{<}1$, the output state shows other photon terms that lead to a mean photon-number $\mathcal{N}_{c,d}{=}p_1\left(1\pm p_0\cos\phi\right)$ at the beamsplitter outputs, {see Supplementary Information}. That is, if states are pure in the photon-number basis, their interference leads to oscillations measured at the output of the interferometer device, with a visibility amplitude equal to the vacuum population $p_0$. 

The previous example describes the idealised case of pure states---instances non-existing in the physical world. To account for impurity  in the photon-number basis, we consider that each light wavepacket impinging on the beamsplitter is described by a density matrix $\rho_{S}{=}\lambda\rho_\text{pure}{+}(1{-}\lambda)\rho_\text{mixed}$, with $\rho_\text{pure}{=}|\Psi_i\rangle\langle\Psi_i|$ a pure state {($i{=}a,b$)}, $\rho_\text{mixed}{=}\text{diag}\{p_0,p_1\}$ a diagonal matrix, and $0{\leq}\lambda{\leq}1$ a parameter tuning the photon-number purity. Moreover, limited purity in the frequency domain is taken into account by the non-unity mean wave-packet overlap $M$ between interfering photons. It can be shown, see Supplementary Information, that such interfering input states result in
	\begin{equation}\label{eq:nsingles}
		n_{c,d}=\frac{1}{2}\left(1\pm v\cos\phi\right),
	\end{equation}
where $n_{c,d}{=}{\mathcal{N}_{c,d}}{/}\left({\mathcal{N}_c{+}\mathcal{N}_d}\right)$ oscillate with a visibility $v{=}\lambda^2p_0\sqrt{M}$. We observe, from Eq.~\ref{eq:nsingles}, that if the interfering states are distinguishable ($M{=}0$), or if the state is emitted in a statistical mixture of photon-numbers ($\lambda{=}0$), then $v$  vanishes. Thus, observing $v{\neq}0$ implies that neither case is true: the state contains quantum coherences in the photon-number basis.

Coherent driving of a two-level system creates a quantum superposition of ground and excited state, with a relative phase governed by the classical phase of the driving laser. If this coherence is transferred to the emitted light state through spontaneous emission, we obtain a photonic state with coherences between the vacuum and one-photon components. We test this hypothesis by performing the above described interferometric measurements. {To do so, we utilise an unbalanced MZI with a path-length difference matching the temporal separation of consecutive emitted wavepackets from the quantum dot to temporally overlap them on {a} beamsplitter, see Figure~\ref{fig:2}a. The free space part of the MZI leads to small path variations in the order of the photon wavelength, acting as} the previously described phase $\phi$.

Figure~\ref{fig:2}b shows our measurements of $n_{c,d}$ for pulse areas $A{=}0.61\pi$, and $A{=}0.14\pi$. {The single detector counts undergo clear oscillations with time, as the optical phase $\phi$  freely evolves in time within the interferometer---evidencing quantum coherence in the photon-number basis}. As predicted, the amplitude of  the oscillations {increases with} the vacuum population, {controlled here} by choosing the driving pulse area. Figure~\ref{fig:2}c shows the extracted oscillation visibilities, obtained from maxima and minima of $n_{c,d}$ with respect to $\phi$, for different values of single-photon countrates (bottom axis) as the pulse area varies within $0{<}A{\le}\pi$ (top axis). We observe the expected increase in visibility when increasing the vacuum part. The visibility $v$ also depends on the {mean wave-packet overlap}  $M$, which {is extracted from} coincidences counts at the MZI output. We measured {${M_{\pi}{=}0.903{\pm}0.008}$} at $\pi$-pulse excitation, a value limited by a small residual phonon sideband~\cite{redPhonon:Grange} since no spectral filtering was used. We can then tune $M$ via the relative photon polarisation, see Fig.~\ref{fig:2}d, and observe that the oscillation visibility vanishes for distinguishable photons, as expected. We observe that $v$ is linear in the single-photon countrates, see Fig.~\ref{fig:2}c---accordingly, proportional to vacuum---from which we deduce an average $\lambda{=}0.965{\pm}0.018$ for all pulse areas up to $\pi$-pulse. The  state purity in the photon-number basis $\mathcal{P}{=}\text{Tr}\left(\rho^2\right)$ is extracted knowing $p_0,p_1$, and $\lambda$. We obtain an average value of {$\mathcal{P}{=}0.968{\pm}0.008$} in the full $[0 {-}\pi]$ pulse area range, see Fig.~\ref{fig:2}e, evidencing the high degree of purity. These states are produced on-demand: for each excitation pulse, the device emits a photon-number superposition, with $p_0{+}p_1{=}1$.

{To support the model described above, we consider the situation where a two-level system, with ground $\ket{g}$ and excited state $\ket{e}$, is coupled to a single spatial mode of the optical field, i.e., a one-dimensional atom~\cite{1dAtom:Auffeves}, a model that has been shown to account well for the system under study~\cite{giesz2016}. We calculate the light-field generated by the QD by solving the Lindblad equation that accounts for the evolution of a two-level system---treating the incoming laser field, the interaction unitary Hamiltonian, as well as the non-unitary dynamics of spontaneous emission and pure dephasing, {see Supplementary Information}. {We obtain a system output state that can be written as the density matrix $\rho_{S}$.} This matrix is time integrated over the whole light pulse, an approach that is valid for excitation pulses well below the spontaneous emission time.}

We theoretically obtain the population $p_1$ (respectively, $p_0$), and coherences $\lambda\sqrt{p_0 p_1}$ of $\rho_{S}$ from parameters within the one-dimensional atom model. Pure dephasing contributes to reducing the mean wave packet overlap of the emitted photons, as well as the populations. The solid blue line in Fig.~\ref{fig:2}e shows the calculated populations $p_1$, and the solid green line the corresponding coherences for the case of maximally pure states ($\lambda{=}1$).

To the best of our knowledge, our observations report for the first time the on-demand {direct generation of highly-pure optical quantum states in the photon-number basis.  Such photon-number quantum superpositions were demonstrated for microwave photons, using quantum feedback with Rydberg atoms~\cite{Sayrin:2011fk}, or through synthesized methods using a superconducting phase qubit~\cite{Hofheinz:2009qv}. Here, the quantum superposition is directly obtained from the spontaneous emission of a quantum emitter. }This is observed not only in the weak-excitation regime~\cite{atature2013}---i.e., elastic scattering---where {the atomic population nearly vanishes}, but also up to population inversion. As a result, by adjusting the excitation pulse area, we can generate quantum superpositions of zero and one photon with controlled populations. We note that our measurements provide information of the purity of the quantum state at the output of the emitter. Imperfect photon extraction from the device, or losses in the optical setup have no impact in the presented interferometric measurements, {see Supplementary Information}. In the next part, we study the case of {even stronger driving, under} $2\pi$-pulse {area}, and report on the generation of a quantum superposition of zero-, one-, and two-photon Fock-states.

\begin{figure*}[htp!]
		\includegraphics[width=.85\textwidth]{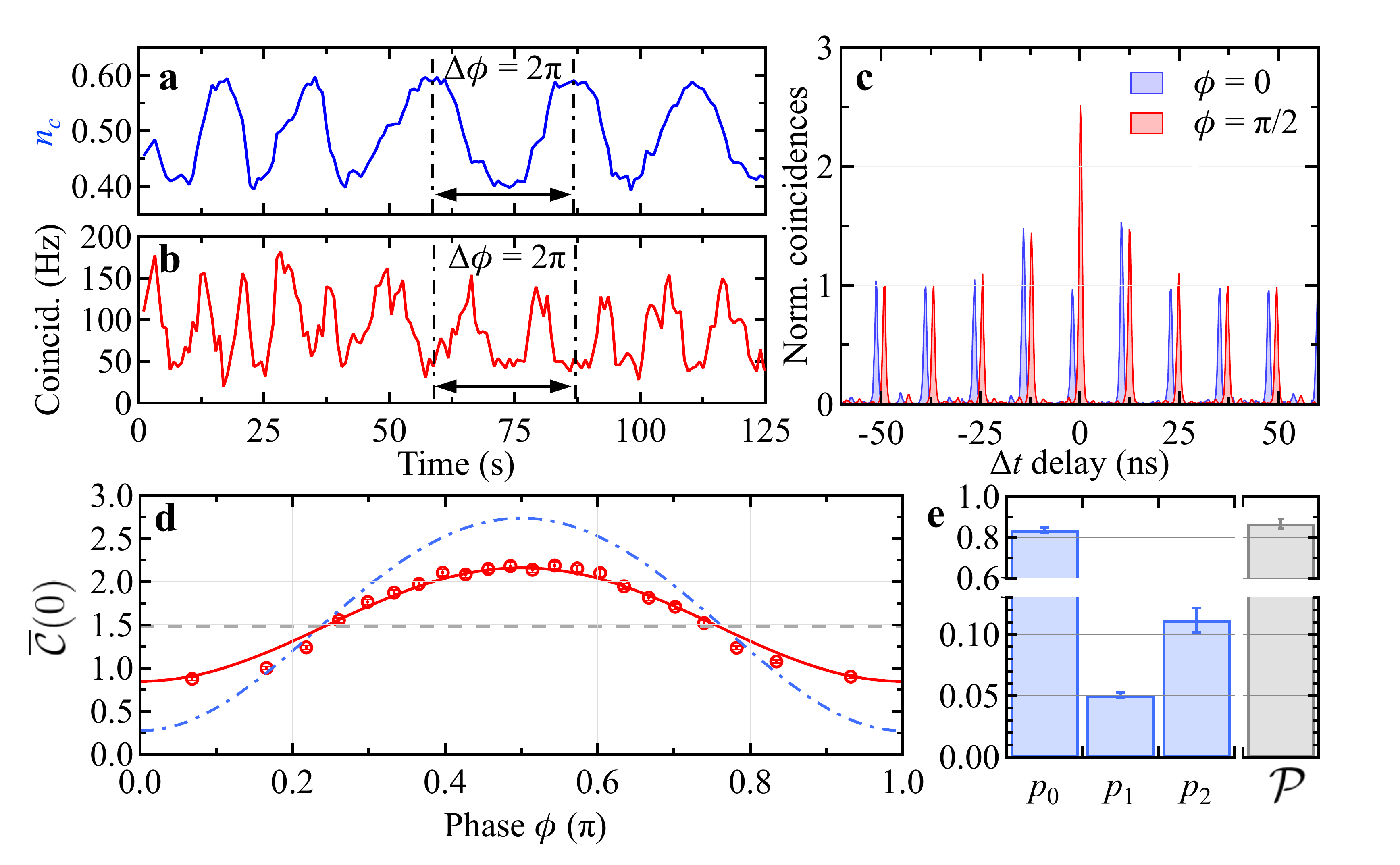}\vspace{-3mm}	
		\caption{{\bf Quantum superposition of vacuum, one, and two photons.} {\bf a} Normalised single counts $n_c$ {as the phase $\phi$ freely evolves in time.} {\bf b} Coincidence rate at zero delay evolving in time {(full data was taken during 1500~s), cycling twice per unit of interferometric phase cycle $\Delta\phi{=}2\pi$---i.e., showing phase super-resolution.} {\bf c} Normalised time-correlated coincidences for a phase $\phi{=}0$ (blue), and $\phi{=}\pi/2$ (red). {\bf d} Normalised coincidences at zero delay $\bar{\mathcal{C}}(0)$ (red circles) as function of $\phi$, with the theoretical prediction (red line) considering the extracted values of $\{p_n,\lambda\}$ described in {\bf e}, and the expected predictions for the same Fock-state populations $\{p_n\}$ in the cases of maximally pure, i.e., $\lambda{=}1$ (blue dashed line), and classical statistical mixtures, i.e., $\lambda{=}0$ (gray dashed line). {\bf e} Populations $\{p_n\}$ and purity $\mathcal{P}$ of the emitted state at $A{=}2\pi$.}
	\vspace{-4mm}	
	\label{fig:3}
	\end{figure*}

\noindent{\bf Quantum superpositions up to two photons~\\}
Strong driving of the atom has been proposed as a mean to generate photon-bundles~\cite{Munoz:2014ty}, and evidences for two-photon emission from an artificial atom has been reported recently by coherently driving a {charged exciton} at $2\pi$-pulse~\cite{Fischer:2017zr}. The excited state population at $2\pi$-pulse drive is expected to be zero, unless some {relaxation} process takes place during the pulse. In particular, as long as the driving pulse duration is not infinitely short, the atom in its excited state shows a non-zero probability to undergo spontaneous emission during the pulse. In such a case, a first photon is spontaneously emitted, and the probability for a second excitation during the pulse is non-zero, leading to the emission of a second photon at the end of the excitation.

The pronounced photon bunching observed at $2\pi$-pulse excitation with our second device QD2, see Fig.~\ref{fig:1}d, quantified by $g^{(2)}_{2\pi}(0){=}2.98{\pm}0.11$, shows that its emission at $A{=}2\pi$ contains non-zero two-photon terms. The generated light wavepacket is then composed by zero-, one- and two-photon Fock-states (higher-number terms are negligible when the excitation pulse-length is significantly shorter than the spontaneous emission time~\cite{Fischer:2017zr}). We now argue that the generated state contains quantum coherences in the photon-number basis. Indeed, the experimental setup depicted in Fig.~\ref{fig:2}a allows to quantify the photon-number populations---including the two-photon component---and degree of purity in this basis.

We learned, from Eq.~\ref{eq:nsingles}, that {the counts of a single detector at the output of the path-unbalanced MZI interferometer {carry} information on the quantum coherence between the zero and one-photon Fock states}. If we now {consider the coincidence counts from the two output detectors as well, we show that we can obtain information on the purity in the number basis up to two photons}. In general, we can extend the previous analysis and consider an input state $\rho_{S}$ now containing terms up to the $|m\rangle$ Fock-state. {We consider a state with the same general form as before, i.e., $\rho_{S}{=}\lambda\rho_\text{pure}{+}(1{-}\lambda)\rho_\text{mixed}$, in a simplified picture where  the state impurity is described through a single parameter $\lambda$, reducing here all coherences in the same way.  Within such framework, } it can be shown, {see Supplementary Information}, that the detected single-click count rates $\mathcal{N}_{c,d}$ at the interferometer outputs $c$ and $d$, see Fig.~\ref{fig:2}a, read
	\begin{equation}
		\mathcal{N}_{c,d}=\frac{1}{2}\left[\langle n\rangle\pm C_{1}\cos(\phi)\right],
		\label{eqG1}
	\end{equation}
where $\langle n\rangle{=}{{\sum_{n}^m}n p_{n}}$ is the system mean photon-number, and $C_{1}{=}\lambda^2{\left({\sum_{n}^m}\sqrt{n p_{n}p_{n-1}}\right)^2}$ is a first-order coherence term, with the summation indices hereafter from $0$ to $m$, {and $\lambda$ accounts for the photon-number purity}. The coincidence rate (at zero delay) follows
	\begin{equation}
	  \mathcal{C}(0)=\frac{1}{8}\left[\langle n(n{-}1)\rangle-C_{2}\cos(2\phi)\right],
		\label{eqG2}
	\end{equation}
with $\langle n(n{-}1)\rangle{=}{\sum_{n}^m n(n{-}1)p_n}$ the non-normalised second-order correlation function, and $C_{2}{=}\lambda^2\left(\sum_{n}^m\sqrt{n(n-1)p_{n}p_{n-2}}\right)^2$ a coherence term of second-order. In virtue of Eqs.~\ref{eqG1} and \ref{eqG2}, we obtain the normalised output coincidences at zero delay $\bar{\mathcal{C}}(0){=}{\mathcal{C}(0)}{/}\left({\mathcal{N}_c\mathcal{N}_d}\right){=}\frac{1}{2}g^{(2)}{(0)}\left(1{-}v_{2}\cos2\phi\right){/}\left(1{-}v_1^2\cos^2\phi\right)$, where $g^{(2)}{(0)}{=}{\langle n(n{-}1)\rangle}{/}{\langle n\rangle^2}$ is the normalised second-order correlation function of the input state, $v_1{=}C_{1}{/}{\langle n\rangle}$ is the single {detector} counts visibility, and $v_2{=}C_{2}{/}{\langle n(n{-}1)\rangle}$ the coincidences visibility. See that Eq.~\ref{eqG2} shows oscillations modulated at twice the phase-dependence of  Eq.~\ref{eqG1}: coherences in the photon-number allow phase super-resolving interferometry.

The state generated at $2\pi$-pulse driving contains up to two-photon terms. In which case we obtain $v_1{=}\lambda^2 p_1\left(p_0{+}2\sqrt{2p_0 p_2}{+}2p_2\right){/}\left(p_1{+}2p_2\right)$, $v_2{=}\lambda^2p_0$, and $g^{(2)}(0){=}2p_2{/}\left(p_1{+}2p_2\right)^2$. Thus, by measuring $v_1$, $v_2$, $g^{(2)}(0)$, and taking into account normalisation $p_0{+}p_1{+}p_2{=}1$, we univocally determine $p_0,p_1,p_2$, and $\lambda$. Figures~\ref{fig:3}a, \ref{fig:3}b show respectively our measurements for $n_c{=}\mathcal{N}_c{/}\langle n\rangle{=}\left(1{+}v_1\cos\phi\right){/}2$, and coincidences proportional to $\mathcal{C}(0){\propto}\left(1{-}v_2\cos(2\phi)\right)$ with a rate specific to the losses of our setup. Figure~\ref{fig:3}c shows our obtained time-correlated coincidence measurements $\bar{\mathcal{C}}{(\Delta t)}$. As predicted, the coincidences at zero delay $\bar{\mathcal{C}}{(0)}$ oscillate with $2\phi$---with minima (maxima) of coincidences occurring for $\phi{=}0$ ($\phi{=}\pi{/}2$). The full phase span for $\bar{\mathcal{C}}{(0)}$ (modulo $\pi$) is shown in Fig.~\ref{fig:3}d. In the case of fully mixed (pure) states in the photon-number basis, i.e., $\lambda{=}0$ ($\lambda{=}1$), oscillation visibilities in the coincidence-counts fully vanish (maximally oscillate), see dashed grey (dot-dashed blue) curve in Fig.~\ref{fig:3}d.

We extract $v_1{=}0.192{\pm}{0.008}$, $v_2{=}0.452{{\pm}{0.038}}$, which together with the measured value of $g^{(2)}_{2\pi}(0){=}2.98{\pm}0.11$, see Fig.~\ref{fig:1}d, and the normalisation of probabilities, results in the distribution $\{p_n,\lambda\}$, with $p_0{=}0.838{\pm}0.012$, $p_1{=}0.051{\pm}0.002$, $p_2{=}0.111{\pm}0.010$, and $\lambda{=}0.734{\pm}0.025$. The generation of light states with $p_2{>} p_1$ is  observed with charged excitons under a strong $2\pi$-pulse drive of the QD, {see Methods}. The state $\rho_S^{2\pi}$ contains zero, one and two photons, with a quantum state purity of {$\mathcal{P}{=}0.870{\pm}0.024$}, see Fig.~\ref{fig:3}(e). Note that the {above theoretical analysis}  for $2\pi$-pulse driving does not {account for the effect of limited  photon indistinguishability ($M{<}1$)}. Accordingly, the reported purity contains both the photon-number purity and indistinguishability imperfections, thus representing a lower bound for the photon-number purity alone.

This state, with $p_2{>}p1$, incidentally resembles other quantum states of interest. The obtained photon distribution $\{p_n^{2\pi}\}$ presents a statistical fidelity $\mathcal{F}^{\text{cat}}{=}\sum_n\sqrt{p_n^{2\pi}p_n^\text{cat}}$ to $\{p_n^\text{cat}\}$, the probability distribution of an even ``Schr\"odinger-cat'' state $|\text{cat}\rangle{\propto}|\alpha\rangle{+}|{-}\alpha\rangle$, with $|\alpha\rangle$ a coherent state, of {$\mathcal{F}^{\text{cat}}{=}0.974{\pm}0.016$} for a small cat state with $|\alpha|^2{=}0.5$. Thus, by simply driving {a charged} quantum dot with $2\pi$-pulses, we are able to generate {other photonic} states that may find applications in coherent-state driven quantum computation~\cite{ralph:cat03,Lund:cat08} and quantum metrology~\cite{milburn:cat04}.

\noindent{\bf Conclusions~\\}
Quantum states with a high degree of purity are essential in all quantum-enhanced technologies. Optical quantum technologies have so far exploited various degrees-of-freedom, such as time{-}frequency, angular momentum, or polarisation~\cite{OBrien:2009rm,Erhard:2018fk}; but not the photon-number due to the absence of suitable sources. Our work demonstrates that state-of-the-art semiconductor QD emitters {not only provide high purity in the frequency basis but also} non-classical photon-number superpositions on-demand. We are now able to generate highly-pure light wave-packets with tuneable zero- and one-photon components. Other non-classical states can be also generated by adjusting the coherent excitation pulse duration and intensity, as shown here driving the atom at $2\pi-$pulse.} We believe that the generation of quantum superpositions of photon-numbers opens new exciting routes for optical quantum technologies. For instance, we now can exploit the interference of these novel photonic states, {potentially impacting on} the complexity of existing quantum-enhanced protocols, such as in quantum computing, or quantum walks.
\\

\noindent{\bf Acknowledgements~} This work was partially supported by the ERC Starting Grant No.~277885 QD-CQED, the French Agence Nationale pour la Recherche (grant ANR SPIQE and USSEPP), the French RENATECH network, a public grant overseen by the French National Research Agency (ANR) as part of the ÔInvestissements d'AvenirÕ programme (Labex NanoSaclay, reference: ANR-10-LABX-0035). J.C.L. and C.A. acknowledge support from Marie Sk\l{}odowska-Curie Individual Fellowships SMUPHOS and SQUAPH, respectively. We thank N. Carlon Zambon for providing technical assistance throughout the project.

\noindent{\bf Author contributions~} The experiments were conducted by J.C.L. and C.A. with help from P.H., C.M. H.O., and L.D.S. The data analysis was done by C.A. and J.C.L. The theoretical modelling was done by A.A., B.R, O.K., C.A., and J.C.L. The cavity devices were fabricated by A.H. and N.S. from samples grown by A.L., and the etching was done by I.S. The project was supervised by L.L., A.A., O.K., and P.S. The manuscript was written by J.C.L., C.A., and P.S., with input from all authors.

\noindent{\bf Competing interests~} N.S. is co-founder, and P.S. is scientific advisor and co-founder, of the single-photon-source company Quandela.
\\

\noindent{\bf\large METHODS~\\}
\noindent{\bf Sample~}
The microcavity samples were grown by molecular beam epitaxy. A $\lambda$-GaAs cavity is
surrounded by a bottom and a top mirror made of 29 and 14 pairs of GaAs/Al$_{0.9}$Ga$_{0.1}As$, respectively. The  mirrors are gradually n-doped and p-doped in order to tune the quantum dot transition through the confined Stark effect. The cavities are centered on the quantum dots using the in-situ optical lithography technique~\cite{dousse2008}. Then the sample is etched and
standard p-contacts are defined on a large frame ($300{\times}300~\mu$m$^2$) connected to the circular frame around the micropillar. A standard n-contact is defined on the sample back surface. A neutral exciton is coupled to the cavity mode for QD1. For the optical measurements, the polarisation of the laser is set so that the fine-structure splitting results to emission in crossed-polarisation, see ref.~\cite{giesz2016}. A positively-charged exciton is coupled to the cavity mode for QD2, and in this case  the circular-polarisation and optical transition rules naturally allows obtaining a signal in the crossed-polarisation configuration. The generation of pure coherent superpositions of zero and one photon has been observed for half a dozen devices, based either on neutral or charged exciton transitions, when exciting below $\pi$-pulse. The generation of light states with $p_2{>} p_1$, on the other hand, is only observed with charged excitons in the present experimental configuration. Indeed, in a crossed-polarisation collection scheme, the neutral exciton spontaneous emission is time delayed by the fine structure splitting~\cite{giesz2016},  preventing an efficient re-excitation within the same pulse.

\noindent{\bf Time-tagged correlation measurements~}
Simultaneous acquisition of single counts and double coincidences are recorded by measuring the photon count-rate and photon event time-tags in the output detectors (Si avalanche photodiodes) of the MZI, which are connected to a computer-controlled HydraHarp 400 autocorrelator. 

Under free evolution of the phase $\phi$ between the two arms of the MZI, the total acquisition time per point has been set to $T_{\textrm{acq}}{=}310$~ms (810~ms) for the results described in Fig.~2 (Fig.~3), with an integration time for the photon time-tags of $T_{\textrm{TT}}{=}200$~ms (500~ms). {Given the relatively fast acquisition of experimental points, the phase $\phi$ remains approximately unchanged during each acquisition run}. The measurement protocol {for each data point} runs as follows: the Hydraharp {autocorrelator} reads the laser clock signal ($24.6700 \pm 0.0026$ ns), a period of time which serves as a reference to determine the photon time tags of the detected  events (accumulated during $T_{\textrm{TT}}$); consecutively, during the interval $T_{\textrm{acq}}{-}T_{\textrm{TT}}$, the countrates in the APDs are averaged, from where the phase $\phi$ is eventually obtained, see Supplementary Information.

\noindent{\bf Data analysis~}
The outcome of the {time-tagged measurements} renders the countrates and two-photon coincidences as function of time (the total integration time is in the order of 10-15 minutes {for a given} pulse area and relative photon polarisation in the MZI). From the oscillation of the single counts, an intensity-to-phase mapping organises the phase-dependent two-photon coincidences as function of the relative phase $\phi$. We use the normalised intensity counts{, e.g., $n_c$,} to assign a corresponding $\phi$ value for {each} given acquisition {time-bin}. For example, at a given {time-bin}, the $n_c$ values that are maximum, minimum, or equal to $n_d$, are mapped to phases $\phi$ equal to $0$, $\pi$, or $\pi{/}2$, respectively. {See Supplementary Information} for further explanations.

\bibliography{qdpure}
\bibliographystyle{naturemag}


\newpage
        
\section{Supplementary Information}
\section{Simulation of the emission process} 
Here we present the model used to compute the quantum state of the emitted field. For the device used in the experiment, the cavity relaxation rate reads $\kappa = 400\pm100$ $\mu$eV while the light-matter coupling constant equals $g=20\pm5$ $\mu$eV (Weak coupling regime). In this case the cavity mode can be adiabatically eliminated and the medium is modeled as a two-level system (TLS) coupled to a waveguide, a 1D atom \cite{1dAtom:Auffeves}. The TLS spontaneous emission rate $\gamma$ accounts for the Purcell enhancement induced by the weak coupling to the cavity. Due to the interaction with the solid-state environment, the TLS also undergoes pure dephasing with a typical rate $\gamma^*$.

The QD is pumped with a pulsed laser, resonant with the QD transition. The pump is modeled by a classical field containing $n_{in}$ photons per pulse on average.
The Hamiltonian of the system in the referential rotating at the pump frequency writes 

\begin{align} \label{H}
\hat{H}=i \hbar \Omega(t)  ( \sigma -  \sigma^\dag)
\end{align}
where $\Omega(t){=}\sqrt{n_{in}\gamma} \xi(t)$ is the classical Rabi frequency and the temporal profile of the pulse $\xi(t)$ is a normalized Gaussian function:
$\xi(t)=\left(\dfrac{4 \ln(2)}{\pi \tau^{2}}\right)^{1/4}\exp(-{t^2}/{\tau^2}8\ln(2))$. The lowering operator is denoted $\sigma{=}\ket{g}\bra{e}$. The dynamics of the QD is ruled by the Lindbladt Master equation:
\begin{equation} \label{L}
\dot{\rho}=-\frac{i}{\hbar}\left[\hat{H},\rho\right]+ D_{\gamma,\sigma}[\rho] + D_{\gamma^*/2,\sigma_z}[\rho]
\end{equation}
where $\sigma_z{=}\left[\sigma^\dag,\sigma\right]$ and $D$ are super-operators defined as
\begin{equation}
D_{\alpha,A}[\rho] = \frac{\alpha}{2}\left(2A\rho A^\dag-A^\dag A\rho-\rho A^\dag A\right) 
\end{equation}

Finally, the polarization of the collected light is orthogonal to the polarization of the the incident laser light, such that the output field $a_{out}$ solely depend on the TLS dipole:
\begin{equation}
 a_{out}=\sqrt{\gamma} \sigma
\end{equation}

The Lindbladt equation is time-integrated, giving access to the total number of emitted photons ${\cal N}_{out}$ during the process and the second-order autocorrelation function $\mathcal{C}(0)$:
 \begin{eqnarray}
 {\cal N}_{out}&=&\int dt \text{Tr}\left[ \rho(t) a_{out}^\dag a_{out} \right]\\
\mathcal{C}(0)&=&
\int dt d\tau \text{Tr}\left[ \tilde{\rho}(t,\tau)  a_{out}^\dag a_{out}\right]
\end{eqnarray}
with:
\begin{equation}
\tilde{\rho}(t,\tau)=U(t,t+\tau) a_{out} \rho(t) a_{out}^\dag U^\dag(t,t+\tau)
\end{equation}
where $U(t_1,t_2)$ is an evolution operator, that satisfies $\rho(t)=U(t_0,t) \rho(t_0) U^\dag(t_0,t)$.

From these parameters we easily deduce the Fock states populations:

\begin{eqnarray}
p_0 = 1-{\cal N}_{out}+\frac{\mathcal{C}(0)}{2}, \\
p_1 = {\cal N}_{out}-\mathcal{C}(0),\\
p_2 = \frac{\mathcal{C}(0)}{2}.
\end{eqnarray}  

The description of the eventual loss of purity induced by pure dephasing would require to model the complete 1D atom (i.e., the TLS coupled to an infinite amount of photonic modes) coupled to a pure dephasing reservoir, which is beyond the scope of the present paper. Instead, we make the assumption that the QD emits pure photonic states $|\psi_{a}\rangle$ and $|\psi_{b}\rangle$ (See main text and below). In this simplified model,  the pure dephasing events mostly impact the time evolution of the TLS population, hence the relative populations of the Fock states. 

Experimental imperfections are modeled in an effective manner. Reduced indistinguishability between the interfering photons due to pure dephasing and/or differences in the temporal profiles of the emitted photons give rise to a reduction of the overlapping factor $M$ (See main text and below). Reduced purity due to decoherence in the Fock space basis is taken into account by the factor $\lambda$ (See main text and below). 
        
\section{Modelling interferences for pure indistinguishable superpositions of Fock-states} \label{sec:model1}  
In this section we present a simple model describing the photon interference taking place in the FBS$_{\textrm{HOM}}$ (defined in the Fig. 2 of the main text) of the MZI between quantum states of light in a coherent superposition in the photon-number basis.  Here we ignore the temporal structure of the field emitted by the QD and treat the input and output ports of the BS as effective modes $a,b,c,d$, respectively. This allows to directly calculate the interference of the two input states, $|\psi_{a}\rangle$ and $|\psi_{b}\rangle$, according to the beam splitter transformations.
	
This basic model allows in principle computing $\mathcal{N}_{c,d}$ and $\mathcal{C}(0)$ between arbitrary photon-number superposition input states \tket{\psi_a} and \tket{\psi_b}, directly interfering in a beam-splitter. The general expression of the two different pure states is $|\psi_{a}\rangle{=}\sum_{n=0}^{m_a} \sqrt{p_n^a} e^{i \alpha_n} \frac{(a^\dagger)^n}{\sqrt{n!}}|0_a\rangle$ and $|\psi_{b}\rangle{=}\sum_{n=0}^{m_b} \sqrt{p_n^b} e^{i \left(\beta_n+n\phi\right)} \frac{(b^\dagger)^n}{\sqrt{n!}}|0_b\rangle$, with $\sum_{n=0}^{m_{a,b}} p_n^{a,b}{=}1$, considering different maximal number of photons $m_{a,b}$ and different relative phases among Fock terms $\alpha_n$ and $\beta_n$ for each input state. Note that the $|\psi_{b}\rangle$ state acquires an additional phase $\phi$, emulating its passage through one of the arm of the MZI. Assuming that the QD is emitting for each laser pulse the same kind of state $|\psi\rangle$, we take $m_a=m_b$ and $\alpha_n{=}\beta_n$.
	
	Under such conditions, we write the initial input state as $|\psi_{\textrm{in}}\rangle{=}|\psi_{a}\rangle \otimes|\psi_{b}\rangle$ and we apply the transformation of a balanced beam splitter on the creation operators $a^\dagger,b^\dagger$ following $a^\dagger{=}\left(c^\dagger{+}d^\dagger \right)/\sqrt{2}$ and $b^\dagger{=}\left(c^\dagger{-}d^\dagger \right)/\sqrt{2}$. The final output state $|\psi_{\textrm{out}}\rangle$ contains all the information to calculate the $\mathcal{N}_{c,d}$ and $\mathcal{C}(0)$. The expressions for single detector counts, simultaneous two detector coincidence counts and visibility of single detector counts read,
	
	\begin{eqnarray}
		\mathcal{N}_{c,d}&=&\sum_{n_{c,d}} n_{c,d} |\langle n_{c,d} | \psi_{\textrm{out}} \rangle|^2 \label{eq:ncdexprs}\\ 
		\mathcal{C}(0)&=&\sum_{n_c, m_d} n_c m_d |\langle n_c m_d | \psi_{\textrm{out}} \rangle|^2 \label{eq:cabexprs} \\ 
		v&=&\max_\phi \frac{\mathcal{N}_c-\mathcal{N}_d}{\mathcal{N}_c+\mathcal{N}_d} \label{eq:v}
	\end{eqnarray}
        
        Then, the normalised simultaneous two detector coincidence counts that we measure after the MZI, $\overline{\mathcal{C}}(0)$, is obtained following $\overline{\mathcal{C}}(0){=}\mathcal{C}(0)/(\mathcal{N}_{c}\mathcal{N}_{d})$. The normalised single counts in detectors $c,d$ will be obtained by simply dividing each expression by the total amount of single detector counts $\mathcal{N}_{c}{+}\mathcal{N}_{d}$. We now restrict the study to the $N{=}1,2$ cases describing our experiments, for an arbitrary relative phase $\alpha_n$ between the Fock terms in each input state.

\subsection{Vacuum and one-photon}
In the first case up to $N{=}1$ photons, the two input states take the form $|\psi_{a}\rangle{=}\left(\sqrt{p_0}{+}\sqrt{p_1} e^{i \alpha_1} a^\dagger \right) |0_a\rangle$ and $|\psi_{b}\rangle{=}\left(\sqrt{p_0}{+}\sqrt{p_1} e^{i \left(\alpha_1{+}\phi\right)} b^\dagger \right) |0_b\rangle$. The expression of the output state resulting from the interference of $|\psi_{a}\rangle$ and $|\psi_{b}\rangle$ is explicitly written in the next equation:
        
         \begin{eqnarray}\label{eq:stateoutp0p1}
		 |\psi_{\textrm{out}}\rangle=p_0|0,0\rangle + \frac{p_1}{\sqrt{2}} e^{i (2 \alpha_1+\phi )} \left(|2,0\rangle-|0,2\rangle\right) \nonumber \\ 
		 +\sqrt{2 p_0 p_1} e^{i (\alpha_1+\frac{ \phi }{2})}\left( \cos \frac{\phi }{2}|1,0\rangle - i \sin\frac{\phi }{2}|0,1\rangle \right)
	\end{eqnarray}  
        
For the sake of clarity, we have removed the explicit indication of the output modes $c$ and $d$ in the Fock-states, so that $|n_c,m_d\rangle {\equiv} |n,m\rangle$. It is worth mentioning that the well known NOON state, $1/\sqrt{2}\left(|2,0\rangle-|0,2\rangle\right)$, is retrieved when $p_0=0$. Instead, if $p_0\neq0$ a term with photon states $|1,0\rangle$ and $|0,1\rangle$ appear (see second line in Eq.~\ref{eq:stateoutp0p1}), which produces the oscillation of single counts as function of the MZI phase $\phi$, reported in Fig.~2 of the main text. In this case, and following Eq.~\ref{eq:ncdexprs}, the visibility of the normalised single detector counts $n_{c,d}$ is $v{=}p_0$, (where obviously $\overline{\mathcal{C}}(0){=}0$) and, as we can see, there is no influence of $\alpha_1$ in the observed oscillation of the single detector counts. For this reason, our measurements of single detector counts and two detector coincidence counts at the output of the MZI do not allow to retrieve information about the $\alpha_1$ phase value of the emitted states of light. 
             
In the particular case of interfering two states of superposition of vacuum and one photon, but with different phases $\alpha_1, \beta_1$, then the normalised single detection counts read $n_{c,d}{=}(1{\pm}p_0 \cos(\alpha_1{-}\beta_1{-}\phi))/2$. Therefore, a finite relative phase $\alpha_1{-}\beta_1$ would be observed as an additional phase in the oscillation of $n_{c,d}$ as function of $\phi$.

\subsection{Vacuum, one, and two photons}         
In a second case regarding up to $N{=}2$ photons, we consider two input states given by $|\psi_{a}\rangle{=}\left(\sqrt{p_0}{+}\sqrt{p_1} e^{i \alpha_1} a^\dagger{+}\sqrt{p_2} e^{i \alpha_2} (a^\dagger)^2/\sqrt{2} \right) |0_a\rangle$ and $|\psi_{b}\rangle{=}\left(\sqrt{p_0}{+}\sqrt{p_1} e^{i \left(\alpha_1 {+} \phi\right)} b^\dagger{+}\sqrt{p_2} e^{i \left(\alpha_2 {+} 2\phi\right)} (b^\dagger)^2/\sqrt{2} \right) |0_b\rangle$, which have identical $\{p_n\}$ distributions. The resulting state $|\psi_{\textrm{out}}\rangle$ of the interference takes a expression given by:       
       \begin{widetext}
	\begin{eqnarray}
		|\psi_{\textrm{out}}\rangle=p_0 |0,0\rangle+\sqrt{2 p_0 p_1} e^{i \left(\alpha_1+\frac{\phi }{2}\right)}  \left( |1,0\rangle\cos \frac{\phi }{2}-i  |0,1\rangle\sin \frac{\phi}{2}\right) +\frac{p_1}{\sqrt{2}}e^{i(2 \alpha_1+\phi )}(|2,0\rangle-|0,2\rangle)\nonumber \\
+ \sqrt{p_0 p_2} e^{i (\alpha_2+\phi )} \left((|0,2\rangle+|2,0\rangle) \cos \phi -i \sqrt{2} |1,1\rangle \sin \phi \right) \nonumber \\
+\sqrt{\frac{p_1 p_2}{2}}e^{i \left(\alpha_1+\alpha_2+\frac{3 \phi }{2}\right)}\left((\sqrt{3} |3,0\rangle-|1,2\rangle) \cos\frac{\phi }{2}+i (\sqrt{3} |0,3\rangle-|2,1\rangle) \sin \frac{\phi }{2}\right)\nonumber \\
+\frac{p_2}{4}e^{i 2(\alpha_2+\phi )}\left(\sqrt{6} |0,4\rangle-2 |2,2\rangle+\sqrt{6} |4,0\rangle\right) \\ \nonumber
	\end{eqnarray}
	\end{widetext}
       
Let us consider first that $\alpha_2{=} 2\alpha_1$; this renders a visibility and double coincidence rate of:
       
       \begin{eqnarray}\label{eqS:v1}
		v_1= p_1\frac{p_0+2p_2+2\sqrt{2p_0 p_2}}{p_1+2p_2}\label{eq:vN2alpha0},\\ \label{eqS:C0}
		\mathcal{C}(0)=\frac{p_2}{4}\left(1-p_0\cos(2\phi)\right). \label{eq:cabN2alpha0} 
	\end{eqnarray}
       
     It is worthy to note that in this case, the oscillation of the single detection counts will be observed ($v_1\neq0$) only if $p_1>0$. The overall behaviour of $v_1$ shows that the increase of vacuum is transformed in an increase of the oscillation visibility. It can be shown that in the particular case of up to two photons, and in the generalised case of up to $m$ photons, one can rewrite ${\cal N}_{c,d}$ and ${\cal C} (0)$ in the form of Eqs.~2,3 in the main text.
       
Finally, let us consider that $\alpha_1$ and $\alpha_2$ have nonzero values in the states with up to $N{=}2$ photons. Now the visibility $v_1$ takes a modified form from that given in Eq. \ref{eq:vN2alpha0} (in particular, note the new term $\cos (2 \alpha_1-\alpha_2)$):
       
        \begin{equation}\label{eq:v1angles}
		v_1^{\alpha}=p_1\frac{p_0+2 p_2+2 \sqrt{2 p_0 p_2} \cos (2 \alpha_1-\alpha_2) }{p_1+2 p_2}
	\end{equation}  
        Equation \ref{eq:v1angles} shows that $\alpha_2 {\neq} 2 \alpha_1$ reduces the visibility. It can be demonstrated that, in this case, $\mathcal{C}(0)$ takes the same form as the one shown in Eq. \ref{eq:cabN2alpha0} under the condition  $\alpha_2 {=} 2 \alpha_1$.

\section{Experimental imperfections}
As mentioned above, two kinds of experimental imperfections are considered: decoherence in the Fock-state basis or in the spectral basis, that are respectively quantified by the coherence damping factor $\lambda{<}1$ and the finite overlap between photonic states $M{<}1$. This finite overlap can be due to spectral decoherence (pure dephasing) or to the emission of wave-packets that are not identical.

\subsection{$0$ + $1$ photons decohering in the Fock states basis}

It is useful to rewrite Eqs.~\ref{eq:ncdexprs} as ${\cal N}_c =\langle \psi_{out}| c^\dagger c| \psi_{out} \rangle$ as a function of the input state. Introducing the unitarity matrix $U$ that transforms the input modes $a,b$ into the output modes $c,d$ of the BS, we have $c := \tilde{a} = U^\dagger a U = (a+b)/\sqrt{2}$ (resp. $d := \tilde{b} = U^\dagger b U = (-a+b)/\sqrt{2}$) and $| \psi_{out} \rangle = U |\psi_{in} \rangle$. Finally,
     \begin{eqnarray}
     {\cal N}_c = \langle \psi_{in}| \tilde{a}^\dagger \tilde{a} | \psi_{in} \rangle \\
      {\cal N}_d = \langle \psi_{in}| \tilde{b}^\dagger \tilde{b} | \psi_{in} \rangle
\end{eqnarray}

Denoting as $|\psi^{a}_{in}\rangle {=} \sqrt{p_0} \ket{0_a} {+} \sqrt{p_1} \ket{1_a}$ (resp. $|\psi^{b}_{in}\rangle {=} \sqrt{p_0} \ket{0_b} {+} \sqrt{p_1}e^{i\phi} \ket{1_b}$) the input state in the $a$ (resp. $b$) input port, the global input state is a product $|\psi_{in}\rangle {=} |\psi^{a}_{in}\rangle{\otimes}|\psi^{b}_{in}\rangle$.  By developing, e.g., the product $\tilde{a}^\dagger \tilde{a}$, the interference terms in ${\cal N}_c$ and ${\cal N}_d$ (i.e., oscillating as a function of $\phi$) are found proportional to ${\cal I} = \langle \psi_{in} | a^\dagger b | \psi_{in} \rangle + cc$, where $cc$ stands for complex conjugate. 

The interference term can thus be rewritten as ${\cal I} =  \langle \psi^{a}_{in} | a | \psi^{a}_{in} \rangle \langle \psi^{b}_{in} |b^\dagger | \psi^{b}_{in}  \rangle  + cc$.
In view of modeling the decoherence in the Fock state basis, we now rewrite $\langle \psi^{o}_{in} | o | \psi^{o}_{in} \rangle {=} \mathrm{Tr}[\rho^o_{in} o]$, where $\mathrm{Tr}$ stands for the trace operation, and $\rho^{o}_{in} {=} |\psi^o_{in} \rangle \langle \psi^o_{in} |$ is the density matrix representing the quantum state in the input port $o {=} a,b$:
     \begin{eqnarray}\nonumber
    \rho^{a}_{in}&=& p_0 | 0_a \rangle \langle 0_a | + p_1 | 1_a \rangle \langle 1_a | + \sqrt{p_0 p_1} \lambda | 0_a \rangle \langle 1_a | \\&+& h.c. \\ \nonumber
    \rho^{b}_{in}&=&p_0 | 0_b \rangle \langle 0_b | + p_1 | 1_b \rangle \langle 1_b | + \sqrt{p_0 p_1} \lambda e^{i\phi} | 0_b \rangle \langle 1_b | \\&+& h.c,
\end{eqnarray}
where $h.c.$ stands for hermitic conjugate.

Let us consider the term $ \mathrm{Tr}[ \rho_{in}^a a]$. It reduces to $\mathrm{Tr}[ \rho_{in}^a a] {=} \langle 1_a | \rho_{in}^a a | 1_a \rangle {=}  \langle 1_a | \rho_{in}^a | 0_a \rangle$ and is damped with respect to the ideal case of a pure superposition by a factor $\lambda$. In the same way, the term $ \mathrm{Tr}[ \rho_{in}^b b]$ is damped by a factor $\lambda$. {The decoherence in the Fock state basis thus leads to a damping of the interference term by a factor $|\lambda|^2$, } such as $v \rightarrow |\lambda|^2 v$. The same reasoning can be extended to superpositions of $0$, $1$ and $2$ photons.
	
\subsection{Distinguishable photons}

		\begin{figure*}[htp]
        		\centering
        		\includegraphics[width=0.65\textwidth]{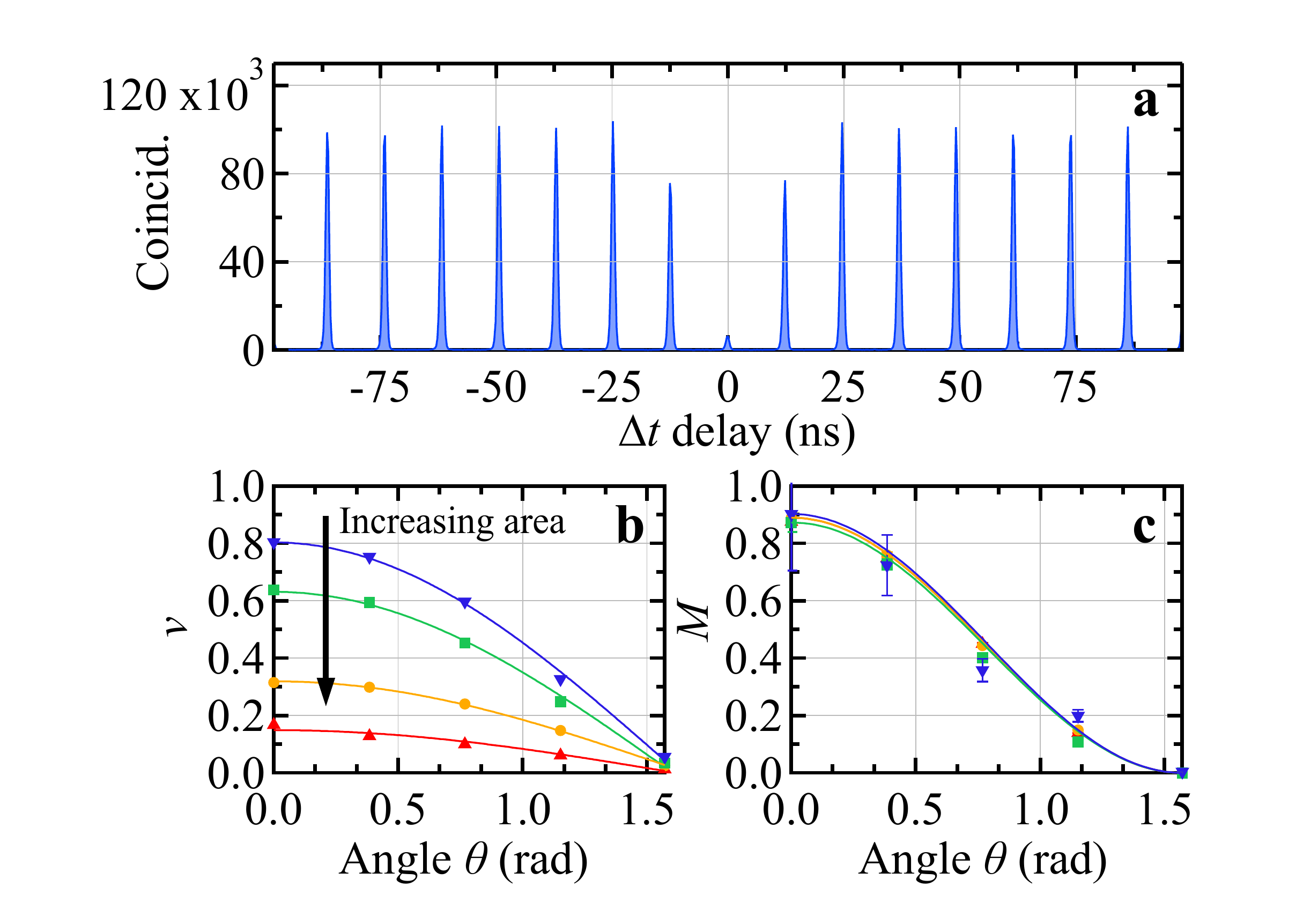}\vspace{-5mm}	
        		\caption{{\bf Probing photon distinguishability.} {\bf a} Indistinguishability, $M{=}0.903{\pm}0.008$, measured in a Hong-Ou-Mandel setup, as explained in the main text, for single-photons emitted with ${\sim}12$~ns separation. {\bf b} Single detector countrate visibility in terms of the polarisation angle $\theta$ of interfering photons, measured for pulse areas $A{=}0.14\pi,0.39\pi,0.61\pi,0.76\pi$. {\bf c} Indistinguishability measured for the same values of $\theta$ and pulse areas used in {\bf b}. Note that, for a given angle $\theta$, the indistinguishability remains the same across all pulse areas. In {\bf b}, {\bf c}, the solid lines are fits to the experimental data according to the equations described in the main text.
		}
        	\label{fig:S1}
        	\end{figure*}
We now model the effect of the distinguishability of the photons impinging on the BS. Supposing that the quantum states sent in $a$ and $b$ are pure but have a temporal structure and correspond to different wave packets. The interference term now reads ${\cal I} {=} \int dt \langle \psi_{in}| a^\dagger(t) b(t)| \psi_{in} \rangle {+} cc$, where we have introduced $o(t)$ the annihilation of a photon at time $t$ in the mode $o$, verifying $[o(t),o^\dagger(t^{'})] {=} \delta(t{-}t^{'})$. One photon in the mode $o$ now writes $| 1_o \rangle {=} \int dt f^*_o(t) o^\dagger |0 \rangle$, where $\int dt |f_o(t)|^2 {=}1$. With these notations, the interference term can be rewritten ${\cal I} {=} \int dt f_a(t) f^*_b(t) + cc$. Denoting $M {=} |\int dt f_a(t) f^*_b(t)|^2$ the real number quantifying the overlap between two photons, we simply find that {the distinguishability of the impinging photons leads to a damping of the interference term by a factor $\sqrt{M}$}, such as $v \rightarrow \sqrt{M} v$. The same effect is reached when the spectral purity of the two photons is reduced by pure dephasing.

In our experiment, we measured a degree of indistinguishability $M{=}0.903{\pm}0.008$, see Fig.~\ref{fig:S1}a, for single-photons emitted ${\sim}12~$ns appart from the QD, and when they are indistinguishable in all other degrees-of-freedom (polarisation, time of arrival) upon impinging at a beamsplitter. From this point, we vary their distinguishability via tuning their relative polarisations. We measured the visibility $v$, as well as the indistinguishability $M$, as function of the angle $\theta$ between the interfering polarisations ($\theta{=}0$ for parallel, and $\theta{=}\pi{/}2$ for orthogonal polarisations) for various pulse areas, see Fig.~\ref{fig:S1}b,c. These measurements are used for displaying the behaviour of $v$ in terms of $M$ shown in Fig.~2d of the main text.


	\section{Role of optical losses}\label{subsec:losses}  

		\begin{figure*}[htp]
        		\centering
        		\includegraphics[width=0.64\textwidth]{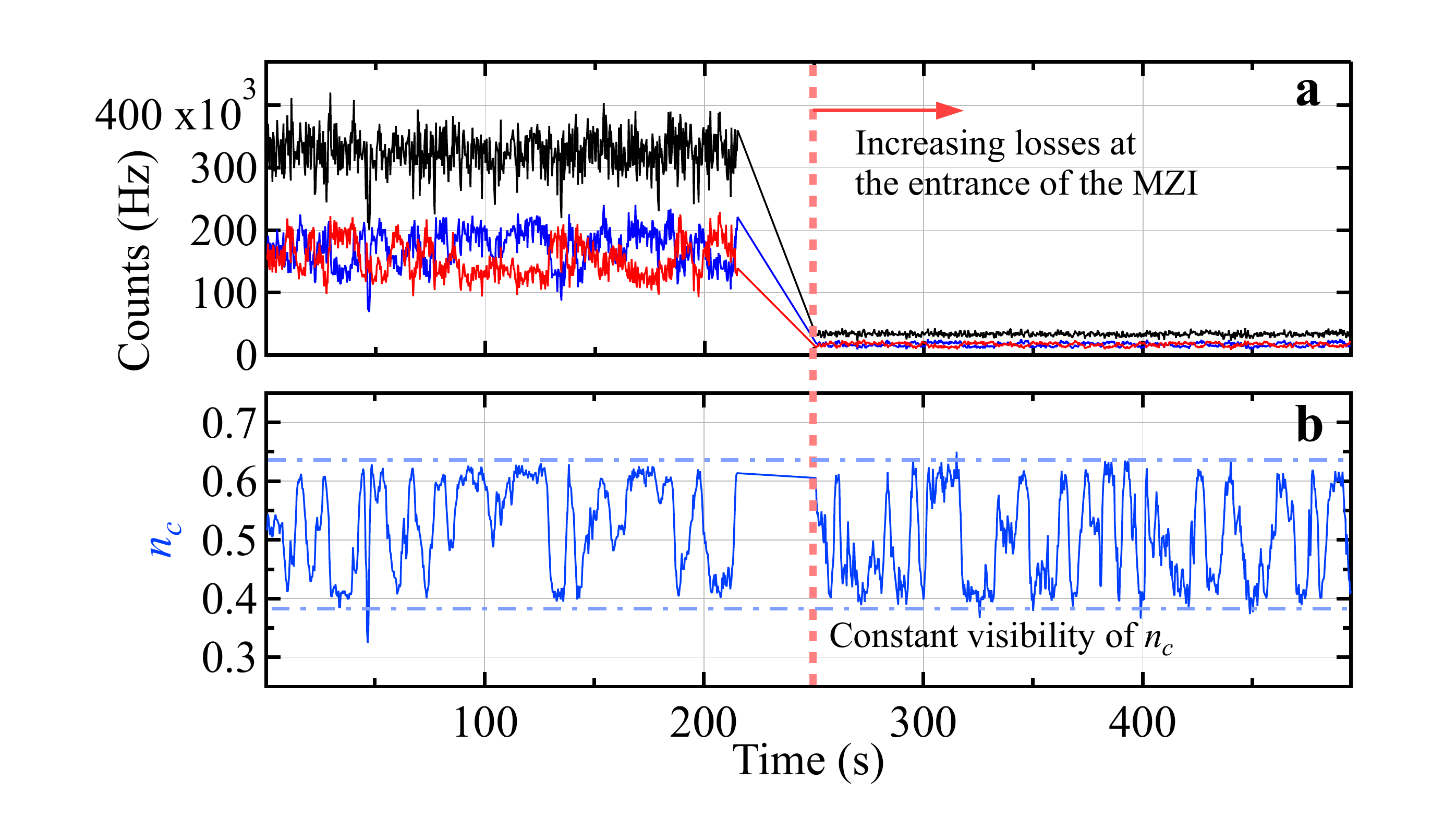}\vspace{-5mm}	
        		\caption{{\bf Role of losses.} {\bf a} Countrates in each detector (red, blue), and their sum (black) as function of time for a pulse area $A{=}0.7\pi$. The countrates evolve in time as they depend on a freely evolving phase $\phi(t)$. {\bf b} Normalised single detector counts ${n}_{c}$, with its oscillation visibility remaining unchanged when introducing extra optical losses.}	
        	\label{fig:S2}
        	\end{figure*}
Here we show that optical losses in the experimental setup have no impact on the observations reported in the main text. To illustrate this, we drive the QD with a pulse area of $A{=}0.7\pi$, and compare measurements with different  amounts of losses.

Figure~\ref{fig:S2}a shows the single detector counts $\mathcal{N}_{c}$ (red) and $\mathcal{N}_{d}$ (blue) measured through our MZI interferometer with its nominal optical transmission, leading to ${\sim}300$kHz of total single-photons (black). At this point, we introduce extra losses to modify the measured countrates for at least one order-of-magnitude, resulting in about ${\sim}10$ times less signal. Figure~\ref{fig:S2}b displays the normalised single detector counts ${n}_{c}$ before and after the extra losses are introduced, thus evidencing that optical losses have no impact on the observed visibilities.

          	\begin{figure*}[htp]
        		\centering
        		\includegraphics[width=.8\textwidth]{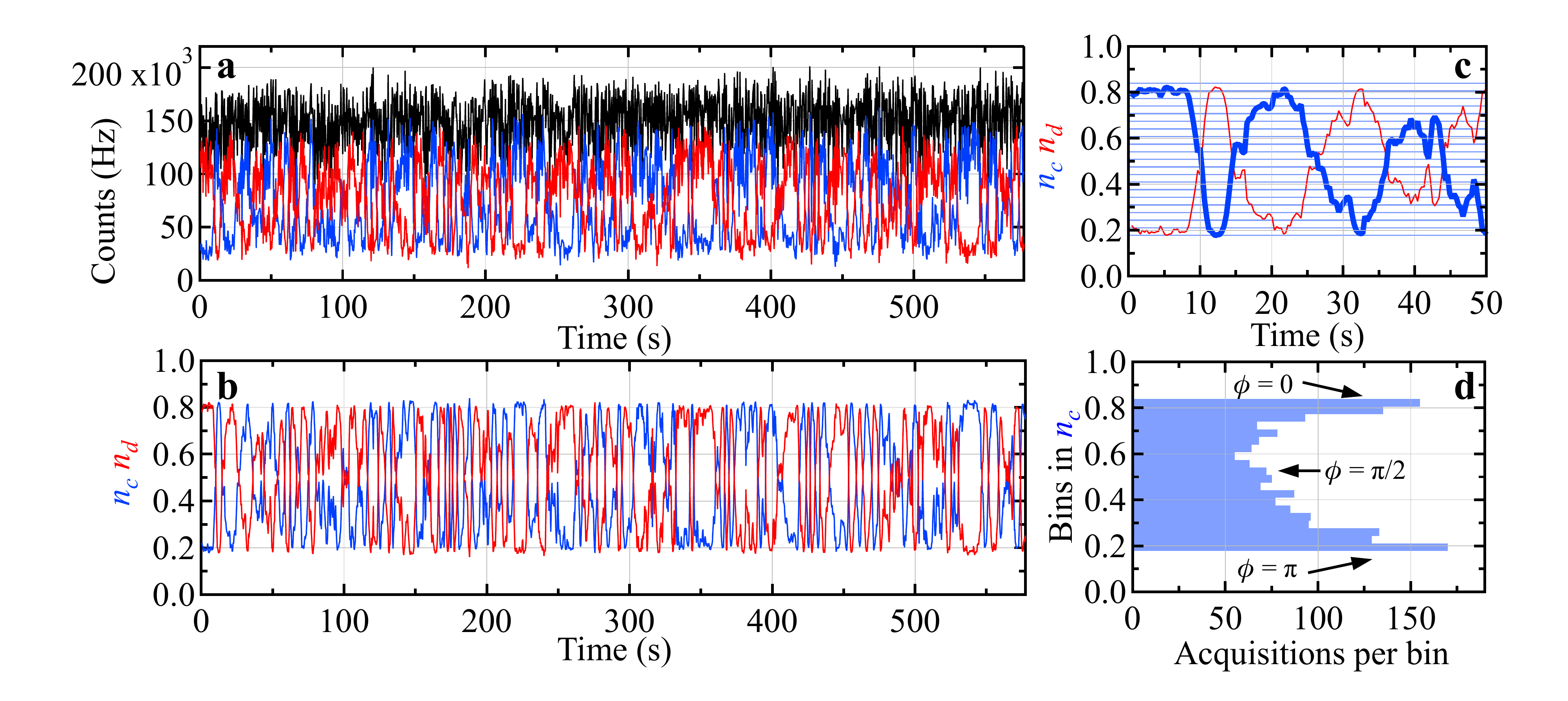}\vspace{-5mm}	
        		\caption{{\bf Retrieving phase values.} {\bf a} Single detector countrates $\mathcal{N}_{c}$ (blue), $\mathcal{N}_{d}$ (red), and their sum (black) for a pulse area $A{=}0.39\pi$, used to obtain the normalised counts ${n}_{c,d}$ shown in {\bf b}. We select 20 intensity bins ({\bf c}), and assign to each a phase value between $\phi{=}0$ and $\phi{=}\pi$ ({\bf d}).
		}	
        	\label{fig:S3}
        	\end{figure*}            

          	\begin{figure*}[htp]
        		\centering
        		\includegraphics[width=.8\textwidth]{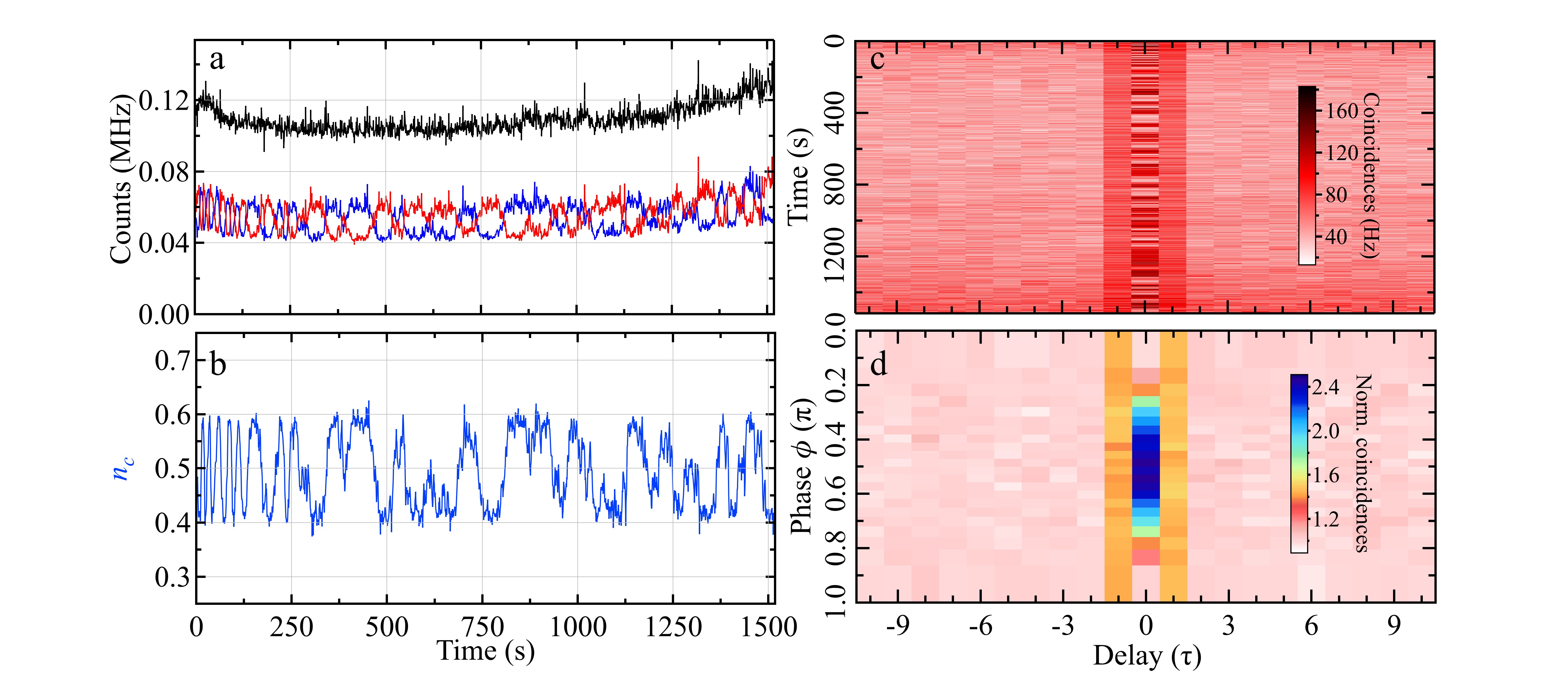}\vspace{-5mm}	
        		\caption{{\bf Phase-resolved coincidence detection.} {\bf a} Single detector countrates (blue, red), and their sum (black) for a pulse area $A{=}2\pi$. {\bf b} The corresponding normalised $\eta_c$. {\bf c} Two-photon countrates accumulated in a 2~ns window for multiple $\tau$ delays between detected events,  evolving during the same time-lapse in {\bf a}. {\bf d} Phase-resolved normalised coincidences obtained from the time-to-phase mapping described in Fig.~\ref{fig:S3}.
		}	
        	\label{fig:S4}
        	\end{figure*} 

From this, we deduce that our measurements contain information on the photonic state prior to any element acting as optical loss, thus at the level of the emitter.


        \section{Time-to-phase mapping}\label{sec:inttophase}        
        
Here we describe how we obtained the interferometric phase $\phi$ from the raw measured countrates freely evolving in time. As an example, we take the measurements of ${n}_{c,d}$ for a pulse area $A{=}0.39\pi$, see Fig.~\ref{fig:S3}a,b. An intensity binning of the data, see Fig.~\ref{fig:S3}c, d, allows to assign a one-to-one correspondence between intensity bins and phase bins, thus obtaining the corresponding values of $\phi$ (modulo $\pi$) at each instant of time.

Similarly, the raw measurements taken at 2$\pi$-pulse area follow the same analysis, see Fig.~\ref{fig:S4}. Here, we monitor single detector, see Fig.~\ref{fig:S4}a,b, and coincidence counts simultaneously, see Fig.~\ref{fig:S4}c. Each time-bin obtained from the single counts is assigned to one phase-bin. Finally, this information is used to obtain the phase-resolved coincidence measurements, see Fig.~\ref{fig:S4}d.

\end{document}